\documentclass[12pt,a4paper,english]{PoS}
\usepackage{lmodern}

\usepackage[T1]{fontenc}
\usepackage[latin9]{inputenc}
\usepackage{float}
\usepackage{amsmath}
\usepackage{amssymb}
\usepackage{graphicx}
\usepackage[numbers]{natbib}

\makeatletter

\pdfpageheight\paperheight
\pdfpagewidth\paperwidth

\title{Determination of Karsch Coefficients for 2-colour QCD}

\ShortTitle{Determination of Karsch Coefficients for 2-colour QCD}

\author{\speaker{Seamus Cotter}\\
	Department of Mathematical Physics, National University of Ireland Maynooth,
	Maynooth, County Kildare, Ireland\\
	E-mail: \email{seamus.cotter@nuim.ie} }

\author{Pietro Giudice\\
	Universit\"at M\"unster, Institut f\"ur Theoretische Physik,
	M\"unster, Germany\\
	E-mail: \email{p.giudice@uni-muenster.de} }

\author{Simon Hands\\
	Department of Physics, College of Science,
  	Swansea University, Swansea, United Kingdom\\
	E-mail: \email{s.hands@swansea.ac.uk} }

\author{Jon-Ivar Skullerud\\
	Department of Mathematical Physics, National University of Ireland Maynooth,
 	Maynooth, County Kildare, Ireland\\
	E-mail: \email{jonivar@thphys.nuim.ie} }

\abstract{We give an update of results from two-colour, two-flavour QCD. Using a Wilson fermion action we calculate thermodynamic quantities as a function of chemical potential $\mu$. Calculating the Karsch Coefficients non-perturbatively gives us access to the derivative method. Compared to our previously published results, we have improved our analysis leading to revised and more accurate estimates for the renormalised energy density, pressure and the trace anomaly.}

\FullConference{31st International Symposium on Lattice Field Theory LATTICE 2013\\
			July 29  August 3, 2013\\
			Mainz, Germany}

\usepackage{changepage}

\usepackage{graphicx}

\makeatother

\usepackage{babel}
\begin{document}

\section{Introduction}

As part of a larger study on two-colour, two-flavour QCD, we use the
derivative method to calculate thermodynamic quantities, in particular
the renormalised energy density. The major stumbling block is the
accurate calculation of the Karsch coefficients \citep{Levkova2006}.
These are defined as the derivative of input parameters with respect
to measured observables. In this case the input parameters are the
gauge coupling $\beta$, the hopping parameter $\kappa$ and the input
gauge and quark anisotropies $\gamma_{g}$ and $\gamma_{q}$ of our
action $S=S_{G}+S_{Q}+S_{J}$. This consists of a non-improved Wilson
gauge and fermion action along with a diquark source action which
serves to lift the low lying eigenmodes of the Dirac operator: 
\begin{align}
S_{G} & =-\frac{\beta}{N_{c}}\left[\frac{1}{\gamma_{g}}\underset{x,i<j}{\sum}\mathrm{ReTr}U_{ij}\left(x\right)+\gamma_{g}\underset{xi}{\sum}\mathrm{ReTr}U_{i0}\left(x\right)\right],\\
S_{Q} & =\underset{x,\alpha}{\sum}\left[\bar{\psi}^{\alpha}\left(x\right)\psi^{\alpha}\left(x\right)+\gamma_{q}\kappa\bar{\psi}^{\alpha}\left(x\right)\left(D_{0}\psi\right)^{\alpha}\left(x\right)\right]+\kappa\underset{x,\alpha,i}{\sum}\bar{\psi}^{\alpha}\left(x\right)\left(D_{i}\psi\right)^{\alpha}\left(x\right),\\
S_{J} & =\kappa j\underset{x}{\sum}\left[\psi^{2tr}\left(x\right)C\gamma_{5}\tau_{2}\psi^{1}\left(x\right)-\bar{\psi}^{1}\left(x\right)C\gamma_{5}\tau_{2}\bar{\psi}^{2tr}\left(x\right)\right].
\end{align}

We define $\beta_{s}=\frac{\beta}{\gamma_{g}}$, $\beta_{t}=\gamma_{g}\beta$,
$\kappa_{s}=\kappa$ and $\kappa_{t}=\gamma_{q}\kappa$. We then calculate
the Karsch coefficients non-perturbatively by measuring the lattice
spacing $a_{s}$, the pion/rho meson mass ratio $M=\frac{m_{\pi}}{m_{\rho}}$,
and the measured gauge $\xi_{g}$ and quark $\xi_{q}$ anisotropies
on several ensembles of anisotropic and isotropic lattices across
a range of values for $\beta$, $\kappa$, $\gamma_{g}$ and $\gamma_{q}$
taken around the central set $\beta=1.9$, $\kappa=0.168$ with $\gamma_{g}=\gamma_{q}=1$,
listed in Table 1. We define the average anisotropy $\xi^{+}=\frac{1}{2}\left(\xi_{g}+\xi_{q}\right)$
and the anisotropy mismatch $\xi^{-}=\frac{1}{2}\left(\xi_{g}-\xi_{q}\right)$,
to ensure we are working along a line of constant physics. As all
thermodynamic quantities are extrapolated to diquark source $j=0$,
the diquark term in the action plays no further role.

Further details about the initial setup are given in an earlier paper
\citep{Cotter2013}. In that paper we overlooked the quark number
density term of the energy density which we include now. A consequence
of the inclusion of this term is that the energy density can be seen
to rely almost totally on this term as the fermionic and gluonic contributions
are small and nearly cancel. As a result the energy density becomes
almost Karsch coefficient independent. This behaviour was conjectured
in an earlier paper \citep{Hands2006c}. We also calculate the pressure
which can be compared to results that used the integral method in
\citep{Cotter2013} and the trace anomaly.

To calculate the spatial lattice spacing $a_{s}$, we use the static
quark potential, and to calculate the gauge anisotropy $\xi_{g}$
we use the sideways potential \citep{Klassen1998a}. For the pion/rho
meson mass ratio $M=\frac{m_{\pi}}{m_{\rho}}$ and the quark anisotropy
$\xi_{q}$ we use the meson dispersion. The results are shown in Table
1 and Figures 1 and 2. Using these measurement results allows for
a four dimensional fit of the measured values for $a_{s}$, $M$,
$\xi^{+}$ and $\xi^{-}$as a function of the input parameters $\beta$,
$\kappa$, $\gamma_{q}$ and $\gamma_{g}$. Inverting the resulting
$4\times4$ matrix gives us the Karsch coefficients shown in Table
2 and 3.

\begin{table}[H]
\noindent \begin{centering}
$\begin{array}{|cccccc|cccc|}
\hline \beta_{s} & \beta_{t} & \kappa_{s} & \kappa_{t} & \gamma_{g} & \gamma_{q} & \xi_{g} & \xi_{q} & M=\frac{m_{\pi}}{m_{\rho}} & a_{s}\left(\textrm{fm}\right)\\
\hline\hline 1.90 & 1.90 & 0.1680 & 0.1680 & 1.0 & 1.0 & 0.968_{-2}^{+2} & 1.035_{-10}^{+8} & 0.798_{-9}^{+4} & 0.178_{-6}^{+4}\\
\hline 2.37 & 1.52 & 0.1680 & 0.1680 & 0.8 & 1.0 & 0.721_{-2}^{+2} & 0.999_{-9}^{+8} & 0.807_{-3}^{+3} & 0.177_{-3}^{+4}\\
1.27 & 2.83 & 0.1680 & 0.1680 & 1.5 & 1.0 & 1.321_{-5}^{+6} & 1.278_{-3}^{+21} & 0.633_{-12}^{+9} & 0.125_{-5}^{+3}\\
1.90 & 1.90 & 0.1800 & 0.1570 & 1.0 & 0.87 & 0.747_{-4}^{+4} & 0.875_{-34}^{+24} & 0.711_{-14}^{+19} & 0.107_{-5}^{+2}\\
1.90 & 1.90 & 0.1470 & 0.1920 & 1.0 & 1.3 & 1.146_{-4}^{+4} & 1.513_{-12}^{+15} & 0.946_{-1}^{+1} & 0.229_{-12}^{+7}\\
\hline 1.80 & 1.80 & 0.1740 & 0.1740 & 1.0 & 1.0 & 0.989_{-3}^{+4} & 1.028_{-14}^{+16} & 0.770_{-6}^{+5} & 0.177_{-7}^{+5}\\
1.90 & 1.90 & 0.1685 & 0.1685 & 1.0 & 1.0 & 0.945_{-5}^{+5} & 1.020_{-11}^{+9} & 0.759_{-13}^{+11} & 0.153_{-18}^{+7}\\
2.00 & 2.00 & 0.1620 & 0.1620 & 1.0 & 1.0 & 0.921_{-5}^{+4} & 0.992_{-9}^{+10} & 0.819_{-6}^{+7} & 0.166_{-2}^{+1}\\
2.00 & 2.00 & 0.1630 & 0.1630 & 1.0 & 1.0 & 0.881_{-5}^{+5} & 1.008_{-6}^{+9} & 0.756_{-7}^{+13} & 0.148_{-1}^{+1}
\\\hline \end{array}$
\par\end{centering}

\noindent \centering{}\caption{Ensemble parameters and measured values for the anisotropies, mass
ratio and lattice spacing.}
\end{table}

\section{Improvements to the determination}

Apart from minor alterations to fit ranges, one area where we immediately
focused our attention was the meson dispersion. Two of the columns
of the $4\times4$ matrix consist of results from the meson dispersion,
any minor improvement could potentially give a large overall improvement.
For the mass fits at zero momentum, this improvement came from a tightening
of the fit range. For the meson dispersion after a similar analysis
and study of the fit ranges an improvement was also seen. On top of
this we also switched from using the continuum definition of the dispersion
relation: 
\begin{eqnarray}
a_{t}^{2}E^{2}\left(p\right) & = & a_{t}^{2}m_{\pi}^{2}+\frac{p^{2}}{a_{s}^{2}\xi_{q}^{2}},\;\;\;\;\;\;\;\;\;\;\textrm{where }p^{2}=p_{x}^{2}+p_{y}^{2}+p_{z}^{2},
\end{eqnarray}
 to the lattice version following \citep{Stamatopoulos2012}: 
\begin{equation}
p^{2}=\frac{4}{a^{2}}\left\{ \sin^{2}\left(\frac{p_{x}a}{2}\right)+\sin^{2}\left(\frac{p_{y}a}{2}\right)+\sin^{2}\left(\frac{p_{x}a}{2}\right)\right\} ,
\end{equation}
This takes into account the discrete values of the momentum on the
lattice. These two improvement resulted in lower error bars across
the board, which can be seen in the latest results for the Karsch
coefficients below. The improved results for the Karsch coefficients
(Table 3) show a reduction in the size of errors from the earlier
determination (Table 2).

\begin{table}[H]
\noindent \begin{centering}
\begin{align*}
\begin{array}{|c|cccc|}
\hline c_{i} & \frac{\partial c_{i}}{\partial\xi_{+}} & a\frac{\partial c_{i}}{\partial a} & M\frac{\partial c_{i}}{\partial M} & \frac{\partial c_{i}}{\partial\xi_{-}}\\
\hline\hline \gamma_{g} & 0.90_{-0.14}^{+0.04} & -0.51{}_{-0.10}^{+0.19} & 0.13_{-0.58}^{+0.32} & 1.4{}_{-1.6}^{+1.2}\\
\gamma_{q} & 0.13{}_{-0.05}^{+0.40} & 0.22_{-0.70}^{+0.12} & -0.55_{-0.29}^{+2.11} & -2.9{}_{-0.6}^{+5.7}\\
\beta & 0.59{}_{-1.37}^{+0.24} & -1.4{}_{-0.5}^{+2.3} & 3.7_{-7.0}^{+1.9} & 8_{-19}^{+8}\\
\kappa & -0.05{}_{-0.02}^{+0.07} & 0.08_{-0.09}^{+0.02} & -0.22{}_{-0.08}^{+0.35} & -0.39_{-0.23}^{+0.88}
\\\hline \end{array}
\end{align*}

\par\end{centering}

\caption{Original Karsch coefficients determination.}
\end{table}

\begin{table}[H]
\begin{align*}
\begin{array}{|c|cccc|}
\hline c_{i} & \frac{\partial c_{i}}{\partial\xi_{+}} & a\frac{\partial c_{i}}{\partial a} & M\frac{\partial c_{i}}{\partial M} & \frac{\partial c_{i}}{\partial\xi_{-}}\\
\hline\hline \gamma_{g} & 0.79{}_{-0.08}^{+0.04} & -0.48{}_{-0.14}^{+0.09} & 0.08{}_{-0.08}^{+0.23} & 1.04{}_{-0.15}^{+0.46}\\
\gamma_{q} & 0.39{}_{-0.03}^{+0.02} & -0.03{}_{-0.04}^{+0.04} & 0.28{}_{-0.08}^{+0.09} & -0.47{}_{-0.15}^{+0.23}\\
\beta & -0.27{}_{-0.19}^{+0.08} & -0.86{}_{-0.36}^{+0.22} & 1.49_{-0.19}^{+0.75} & 1.94{}_{-0.29}^{+1.69}\\
\kappa & -0.01{}_{-0.01}^{+0.01} & 0.05{}_{-0.02}^{+0.02} & -0.12{}_{-0.05}^{+0.14} & -0.11{}_{-0.11}^{+0.02}
\\\hline \end{array}
\end{align*}

\caption{Improved Karsch coefficients determination.}
\end{table}

\begin{figure}[H]
\noindent \raggedright{}\begin{adjustwidth*}{0mm}{-5mm}\includegraphics[scale=0.29]{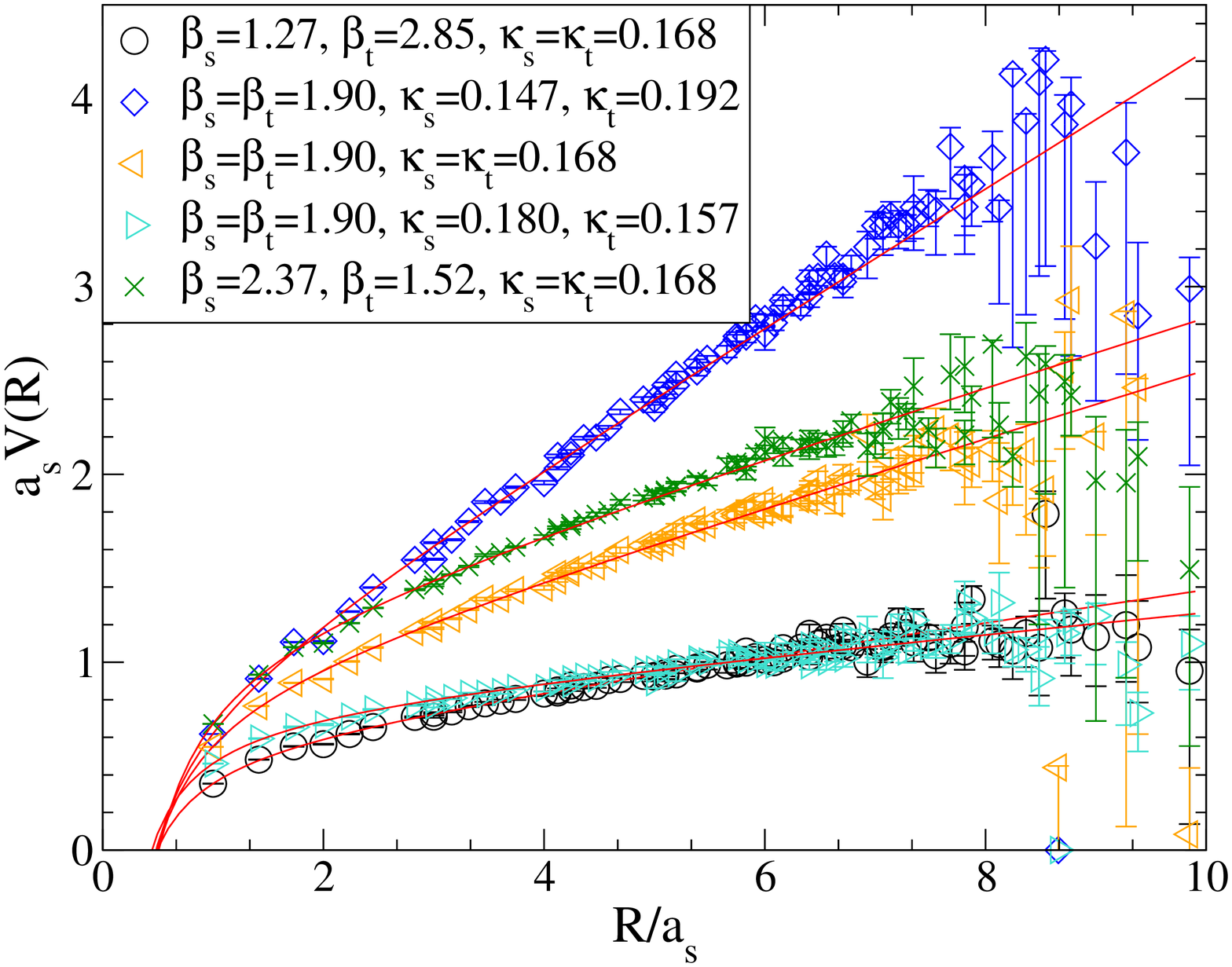}\includegraphics[scale=0.29]{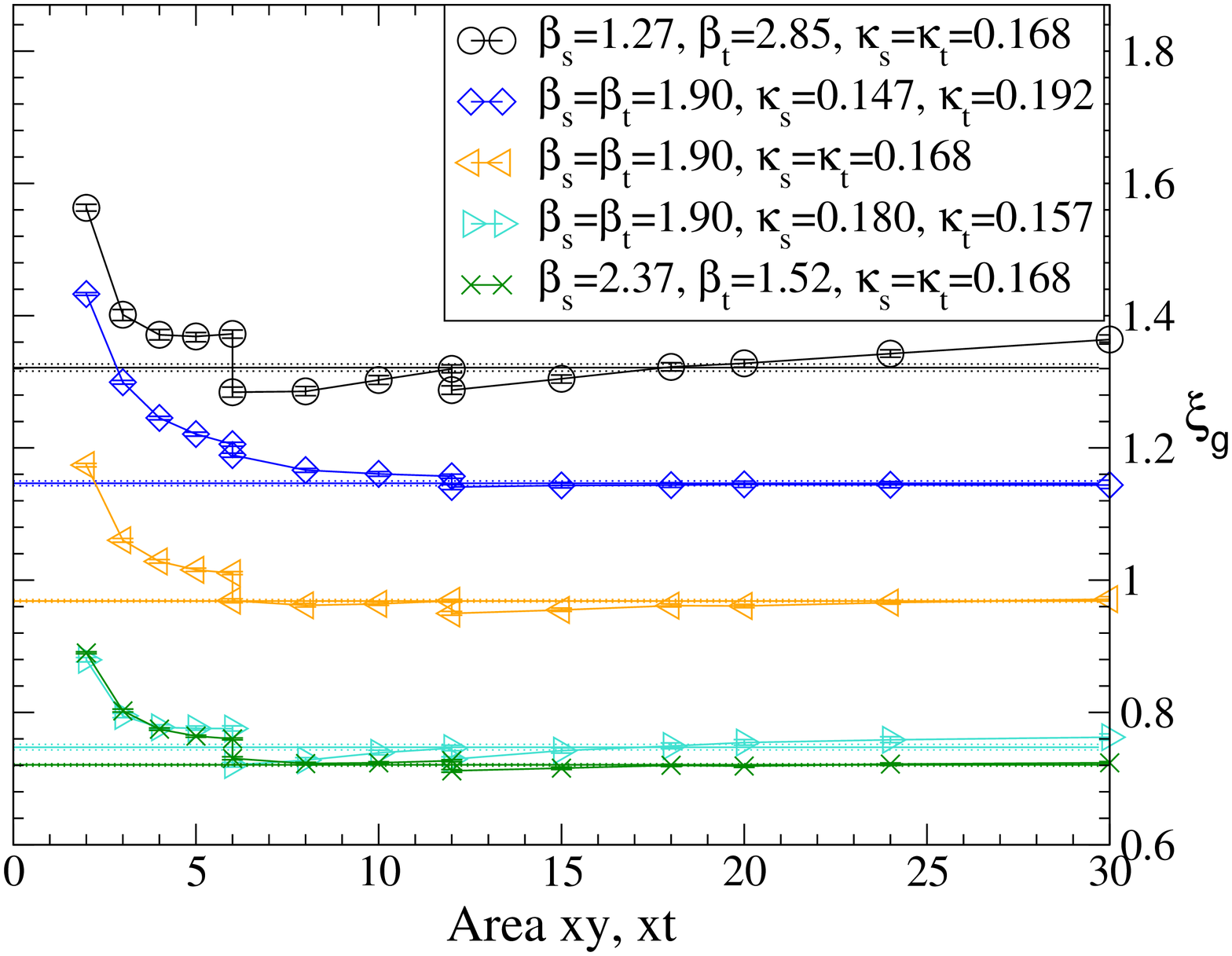}\end{adjustwidth*}\caption{Static quark potential (left) and Sideways potential results (right)
are shown for the central set and the anisotropic sets.}
\end{figure}

\begin{figure}[H]
\noindent \begin{raggedright}
\begin{adjustwidth*}{0mm}{-5mm}\includegraphics[scale=0.29]{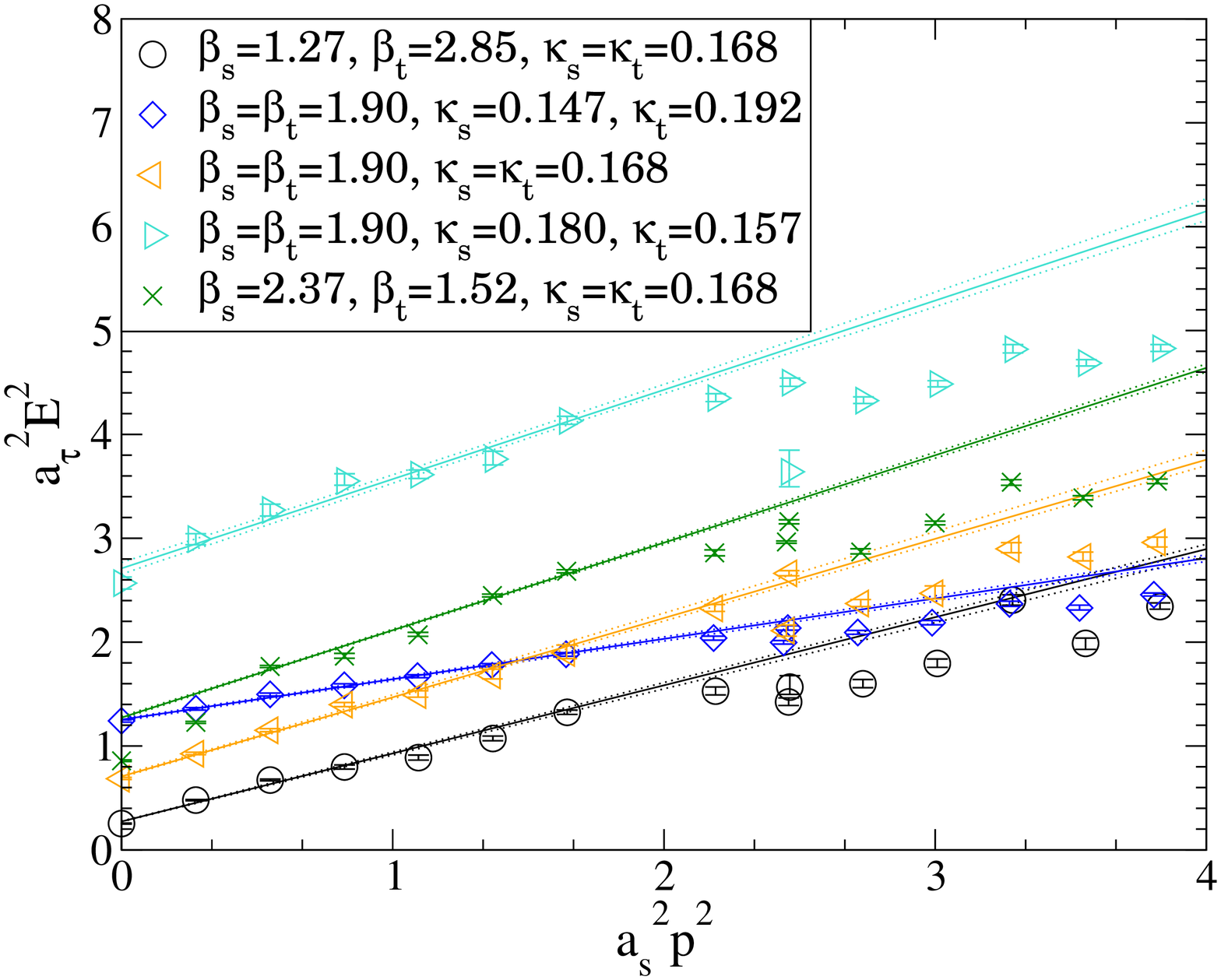}\includegraphics[scale=0.29]{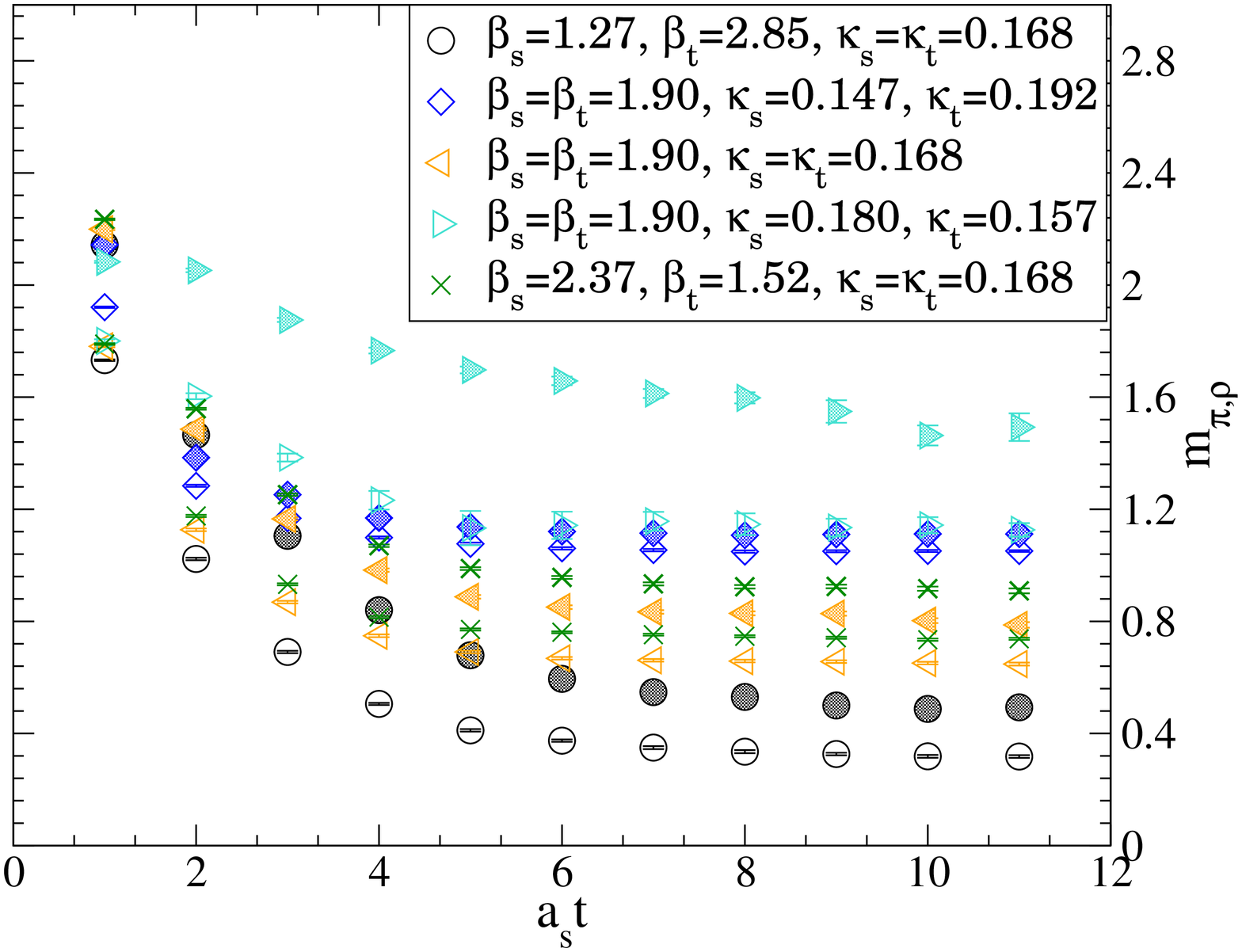}\end{adjustwidth*}
\par\end{raggedright}

\caption{The pion dispersion (left) and the effective mass (right) of the pion
(empty) and rho (shaded) mesons for the central set and the anisotropic
sets.}
\end{figure}

\section{Equation of State}

The partial derivatives must be taken with all other physical parameters
fixed, which means the physical quark mass, and therefore the mass
ratio $M=\frac{m_{\pi}}{m_{\rho}}$ are kept fixed. The energy density
can be derived using the standard thermodynamic relations:
\begin{equation}
\Omega=E-T\mathbb{S}-\mu_{q}N_{q}=-pV=-T\ln\mathcal{Z},
\end{equation}

\begin{equation}
p=\left.\frac{\partial\left(T\ln\mathcal{Z}\right)}{\partial V}\right|_{T},\;\;\;\;\;\varepsilon=\left.\frac{E}{V}\right|_{V},\mathbb{S}=\left.\frac{\partial\left(T\ln\mathcal{Z}\right)}{\partial T}\right|_{V},\;\;\;\;\; n_{q}=\frac{N_{q}}{V}=\frac{1}{V}\frac{\partial\left(T\ln\mathcal{Z}\right)}{\partial\mu},
\end{equation}

which gives us\begin{adjustwidth*}{-5mm}{0mm}
\begin{align}
\varepsilon & =\varepsilon_{g}+\varepsilon_{q}+\mu_{q}n_{q}=-\frac{T}{V}\left\langle \xi\frac{\partial S}{\partial\xi}\right\rangle +\mu_{q}n_{q},\nonumber \\
 & =\frac{3}{N_{s}^{3}a_{s}^{3}N_{t}a_{t}N_{c}}\left[\frac{\beta}{\gamma_{g}}\left\langle \Box_{s}\right\rangle \left(\frac{1}{\beta}\frac{\partial\beta}{\partial\xi^{+}}-\frac{1}{\gamma_{g}}\frac{\partial\gamma_{g}}{\partial\xi^{+}}\right)+\beta\gamma_{g}\left\langle \Box_{t}\right\rangle \left(\frac{1}{\beta}\frac{\partial\beta}{\partial\xi^{+}}+\frac{1}{\gamma_{g}}\frac{\partial\gamma_{g}}{\partial\xi^{+}}\right)\right]\nonumber \\
 & -\frac{1}{N_{s}^{3}a_{s}^{3}N_{t}a_{t}}\left[\gamma_{q}\kappa\left(\frac{1}{\gamma_{q}}\frac{\partial\gamma_{q}}{\partial\xi^{+}}\right)\left\langle \bar{\psi}D_{0}\psi\right\rangle -\kappa\left(\frac{1}{\kappa}\frac{\partial\kappa}{\partial\xi^{+}}\right)\left(4N_{c}N_{f}+\left\langle \bar{\psi}\psi\right\rangle \right)\right]+\mu_{q}n_{q}.
\end{align}

\end{adjustwidth*}

The trace anomaly and pressure follow a similar procedure using the
respective Karsch coefficients ( or $\beta$ functions) :
\begin{eqnarray}
\epsilon-3p & = & \frac{T}{V}\left\langle a\frac{\partial S}{\partial a}\right\rangle ,\;\;\;\;\; p=-\frac{T}{3V}\left[\left\langle a\frac{\partial S}{\partial a}\right\rangle +\left\langle \xi\frac{\partial S}{\partial\xi}\right\rangle \right]+\frac{\mu n_{q}}{3}.
\end{eqnarray}

The angled brackets are vacuum subtracted using results from an ensemble
with volume $16^{3}\times24$, $ja=0.0$ and $\mu=0.0$. We calculate
the energy density, quark number density, trace anomaly and pressure
on 3 volumes $12^{3}\times24$, $16^{3}\times12$ and $16^{3}\times8$,
which translate to $47$MeV, $94$MeV and $141$MeV respectively.
On all three volumes we measure the thermodynamic quantity in question
at diquark source $ja=0.04$ and $ja=0.02$ and extrapolate to zero.

For the energy density, trace anomaly and the pressure we also repeated
the analysis with 100 bootstrap sample values from the Karsch coefficient
determination to estimate their uncertainty, shown as shaded symbols
and dashed error bars. The quark number density (Fig 3, left) is shown
to highlight the dominance of the quark number density term in both
the energy density and the pressure. The energy density (Fig 3, right)
is seen to be almost oblivious to the error coming from the Karsch
coefficient determination except for small $\mu$. The quark number
density shown in Figure 3 is normalised by $n_{SB}^{cont}$ and to
allow for comparison with a previous calculation using the integral
method (semi-filled symbols) \citep{Cotter2013}, the pressure shown
in Figure 4 is normalised by $p_{SB}^{cont}$:
\begin{equation}
n_{SB}^{cont}=N_{f}N_{c}\left(\frac{\mu T^{2}}{3}+\frac{\mu^{3}}{3\pi^{2}}\right),\;\;\;\;\; p_{SB}^{cont}=\frac{N_{f}N_{c}}{12\pi^{2}}\left(\mu^{4}+2\pi^{2}\mu^{2}T^{2}+\frac{7\pi^{4}}{15}T^{4}\right).
\end{equation}

\begin{figure}[H]
\begin{adjustwidth*}{-5mm}{0mm}\includegraphics[scale=0.29]{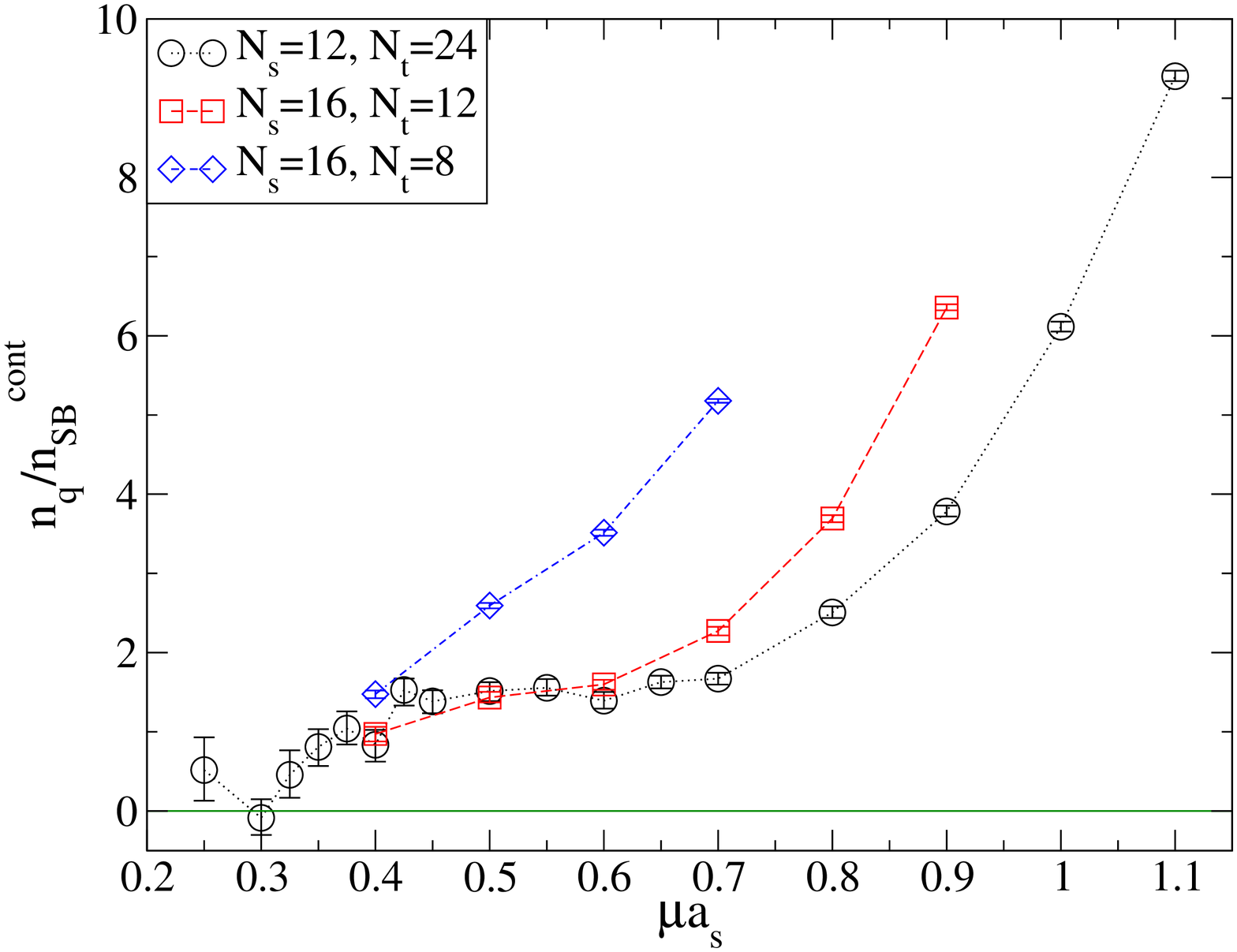}\includegraphics[scale=0.29]{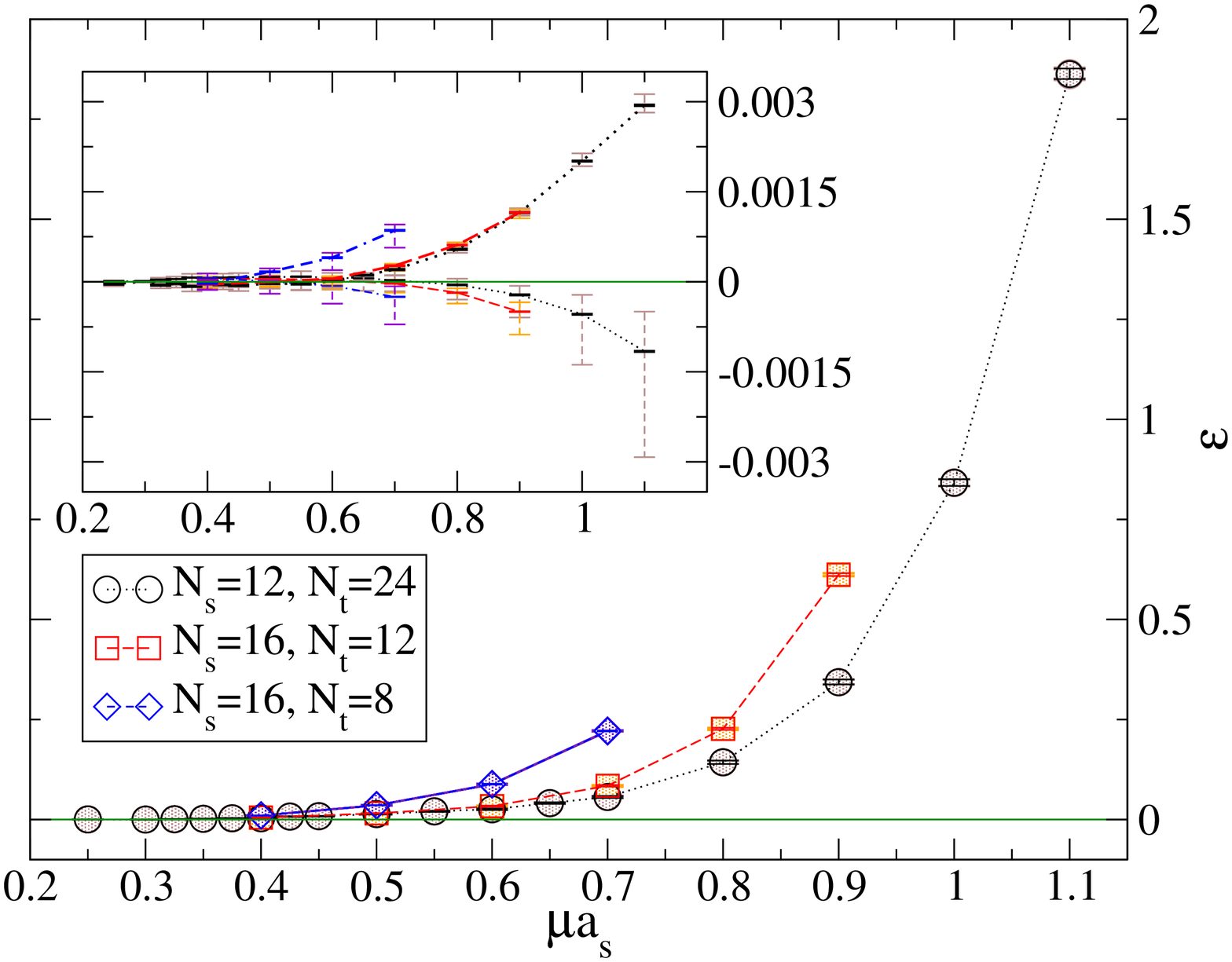}\end{adjustwidth*}

\caption{Quark number density (left), and total energy density (right) as a
function of chemical potential $\mu$. Inset is the fermionic (bold)
and gluonic contributions which come with a Karsch coefficient prefactor.
The shaded symbols in the main plot and the dashed error bars in the
inset denote the Karsch coefficient determination uncertainty.}
\end{figure}

\begin{figure}[H]
\begin{adjustwidth*}{0mm}{-5mm}\includegraphics[scale=0.29]{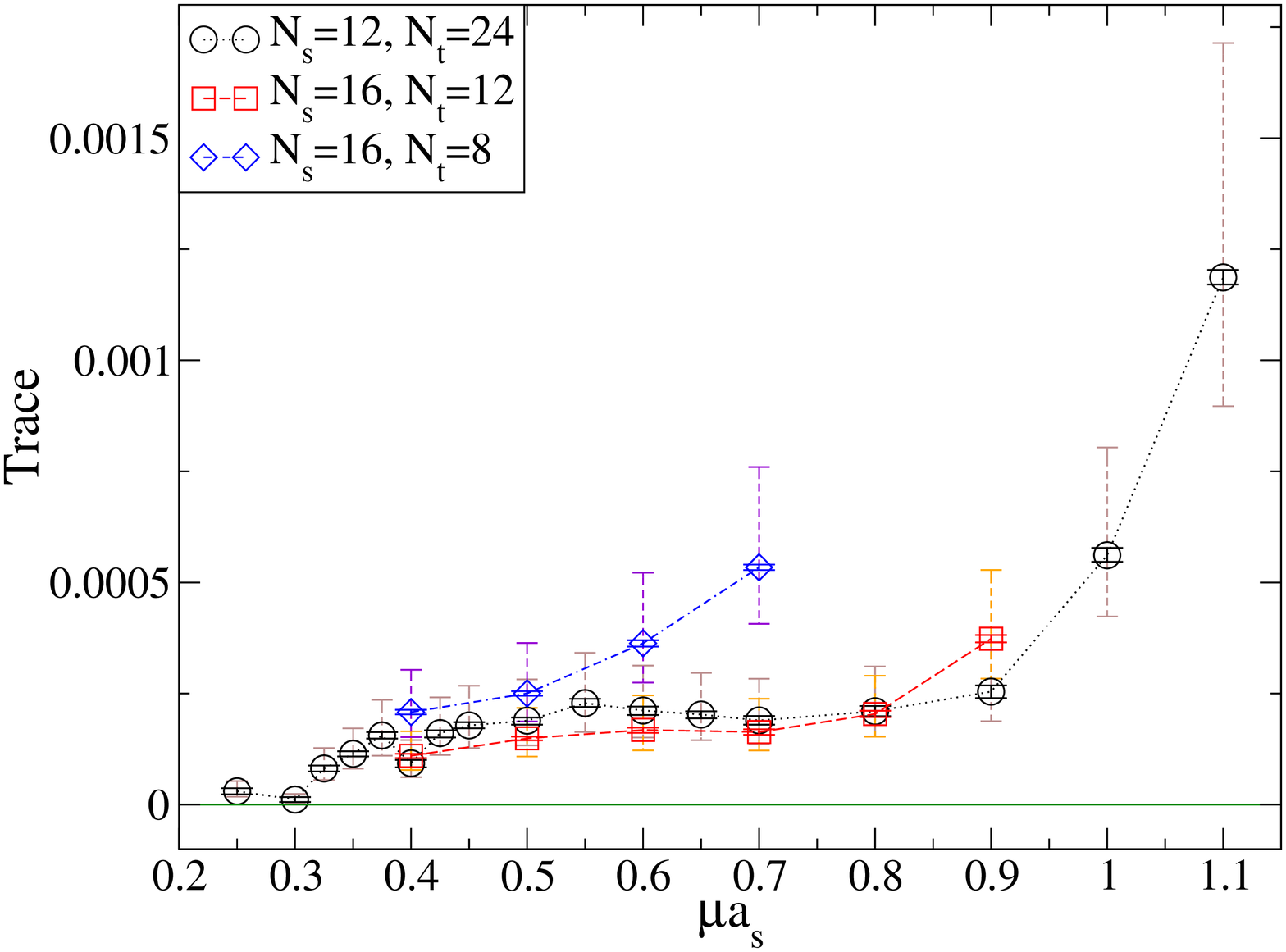}\includegraphics[scale=0.29]{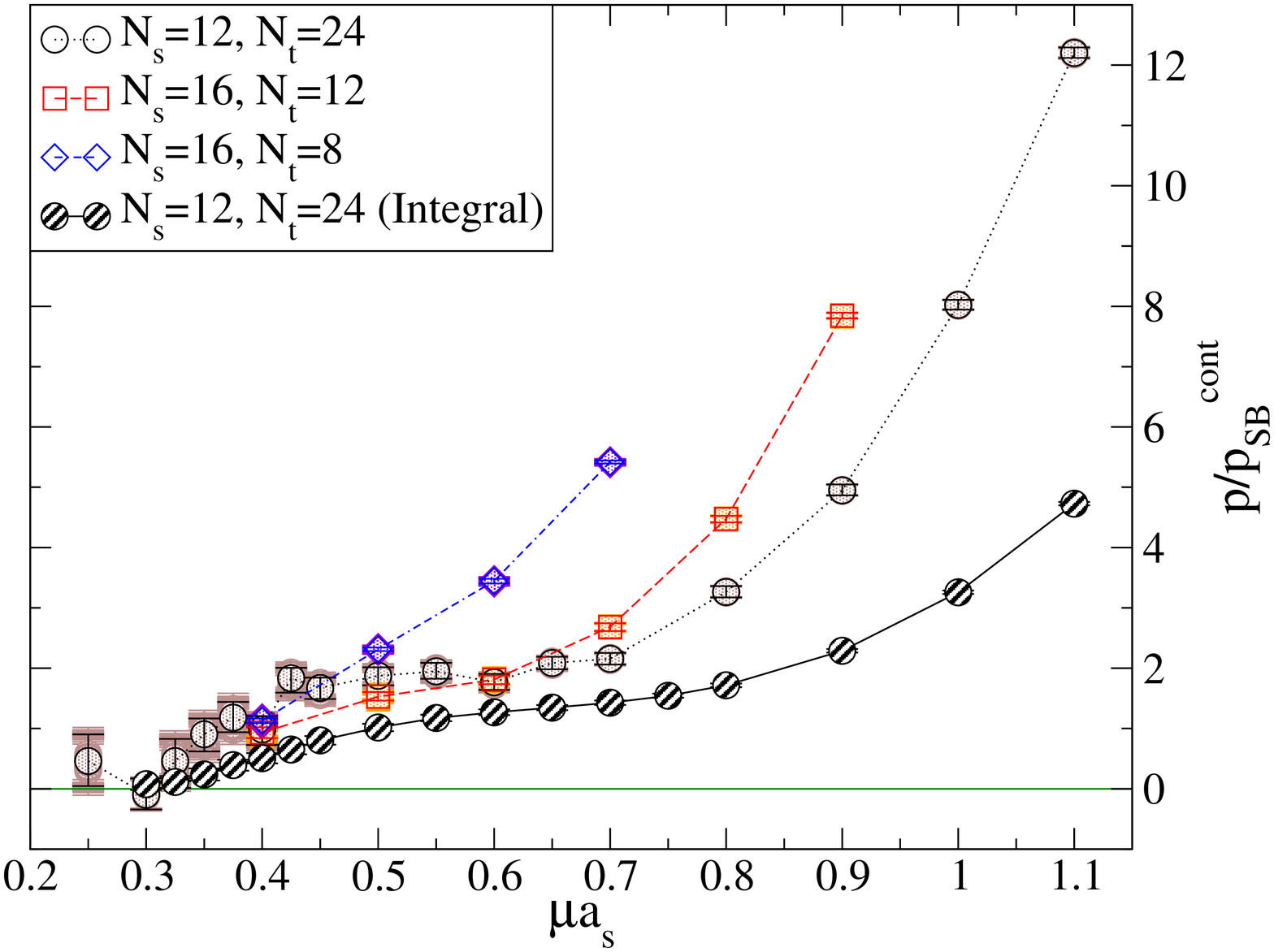}\end{adjustwidth*}

\caption{Trace Anomaly (left) and pressure (right) as a function of chemical
potential $\mu$. The shaded symbols and dashed error bars denote
the uncertainty coming from the Karsch coefficients determination.
For the pressure, the results calculated using the integral method
are also shown (semi-filled symbols).}
\end{figure}

The trace anomaly (Fig 4, left) remains positive, slowly rising at
large $\mu$. As the trace anomaly is closer to zero in size, the
uncertainty in the Karsch coefficients is more apparent. The pressure
(Fig 4, right) is somewhat more sensitive to the difference in values
of the Karsch coefficients at small $\mu$. The values generated on
the $12^{3}\times24$ volume, agree well with the results from the
integral method. As the quark number density does not require the
Karsch coefficients it can be seen that at higher chemical potential
$\mu$, both the pressure and energy density are effectively Karsch
coefficient independent.

\section{Summary and Outlook}

This study demonstrates the feasibility of the derivative method using
non-perturbatively determined Karsch coefficients. Several improvements
are still possible. Our current static potential code doesn't scale
well, and with new ensembles on larger volumes with finer lattice
spacings coming online soon, a newer more efficient plan of attack
is needed. At the moment that looks to be the $W_{0}$ scale from
the Wilson flow \citep{Borsanyi2012b} which we are currently working
on changing to handle $SU\left(2\right)$ configurations rather than
the $SU\left(3\right)$ ones it was designed for. This would also
in principle replace the sideways potential code with which we used
to calculate $\xi_{g}$, although possessing alternative methods and
codes to measure the same quantity allows to control systematic uncertainties.

\section*{Acknowledgments}

This project was part of the UKQCD collaboration and the DiRAC Facility
jointly funded by STFC, the Large Facilities Capital Fund of BIS and
Swansea University. We thank the DEISA Consortium, funded through
the EU FP7 project RI222919, for support within the DEISA Extreme
Computing Initiative. The simulation code was adapted with the help
of Edinburgh Parallel Computing Centre funded by a grant from EPSRC.
JIS and SC acknowledge the support of Science Foundation Ireland grants
08-RFP-PHY1462, 11-RFP.1-PHY3193 and 11-RFP.1-PHY3193-STTF-1.

\bibliographystyle{jphysicsB}
\bibliography{/home/seamus/Documents/link_to_documents/Biblio/biblio}

\end{document}